\documentclass{article}

\usepackage{arxiv}

\usepackage[utf8]{inputenc} % allow utf-8 input
\usepackage[T1]{fontenc}    % use 8-bit T1 fonts
\usepackage{hyperref}       % hyperlinks
\usepackage{url}            % simple URL typesetting
\usepackage{booktabs}       % professional-quality tables
\usepackage{amsfonts}       % blackboard math symbols
\usepackage{nicefrac}       % compact symbols for 1/2, etc.
\usepackage{microtype}      % microtypography
\usepackage{graphicx}
 \usepackage[numbers]{natbib}
\usepackage{doi}
\usepackage{amsmath}
\usepackage{subcaption}
\usepackage{floatrow}
\usepackage{cleveref}       % smart cross-referencing
\newcommand{\svae}{$\mathcal{S}$-VAE }
\newcommand{\nvae}{VAE }
\newcommand{\sae}{$\mathcal{S}$-AE }
\newcommand{\nae}{AE }

\title{Modeling Barrett's Esophagus Progression using Geometric Variational Autoencoders}

\usepackage{authblk}

\author[1]{Vivien van Veldhuizen}
\author[1]{Sharvaree Vadgama}
\author[2]{Onno de Boer}
\author[2]{Sybren Meijer}
\author[1]{Erik J. Bekkers}
\affil[1]{University of Amsterdam, Amsterdam, the Netherlands}
\affil[2]{Amsterdam University Medical Centres, Amsterdam, the Netherlands}
\renewcommand{\shorttitle}{Modeling BE Progression using Geometric VAEs}

\hypersetup{
pdftitle={Modeling Barrett's Esophagus Progression using Geometric Variational Autoencoders},
pdfsubject={cs.LG, q-bio.QM},
pdfauthor={Vivien van Veldhuizen, Sharvaree Vadgama, Onno J. de Boer, Sybren Meijer, Erik J. Bekkers},
pdfkeywords={Oncology, Pathology, Variational Autoencoders, Geometric Deep Learning, Equivariance, Representation Learning},
}

\begin{document}
\maketitle

\begin{abstract}
Early detection of Barrett's Esophagus (BE), the only known precursor to Esophageal adenocarcinoma (EAC), is crucial for effectively preventing and treating esophageal cancer. In this work, we investigate the potential of geometric Variational Autoencoders (VAEs) to learn a meaningful latent representation that captures the progression of BE. We show that hyperspherical VAE (\svae) and Kendall Shape VAE show improved classification accuracy, reconstruction loss, and generative capacity. Additionally, we present a novel autoencoder architecture that can generate qualitative images without the need for a variational framework while retaining the benefits of an autoencoder, such as improved stability and reconstruction quality.
\end{abstract}


\section{Introduction}
\begin{figure}
    \centering\includegraphics[width=1\textwidth]{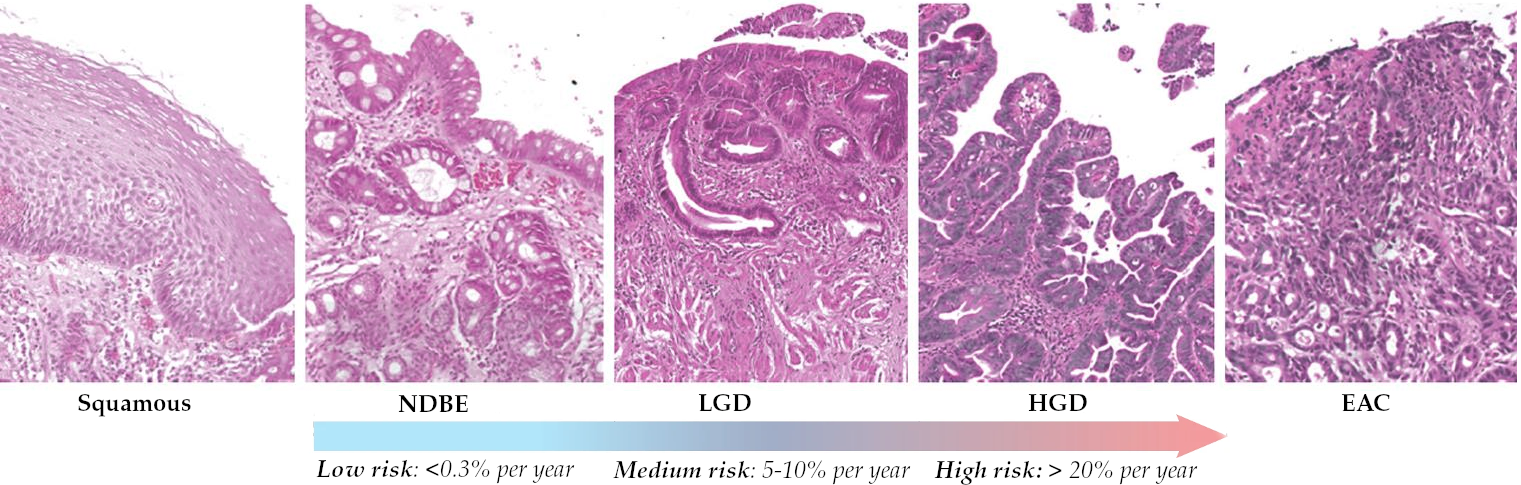}
    \vspace{-6.5mm}
    \caption{Different stages of progression: regular squamous epithelium, non-dysplastic BE, low-grade dysplastic BE, high-grade dysplastic BE, and EAC.\vspace{-4mm}}\label{fig:progression}
\end{figure}

Esophageal adenocarcinoma (EAC) is an aggressive type of cancer with a generally poor prognosis that could benefit from recent advances in machine learning, as it is often diagnosed at a late stage. The only known precursor to EAC, Barrett's Esophagus (BE), progresses through different stages \cite{vanderwelBarrett} (Fig. \ref{fig:progression}), providing an opportunity for early detection and prevention. Currently, the detection of dysplasia relies on subjective assessment by pathologists. Advancements in deep learning have introduced the concept of a \textit{digital pathologist} using convolutional neural networks \cite{de2018survey,barretcnn1}. However, while these models have shown promise, they are limited by a high degree of interobserver variability in labeled training data \cite{bolero}.

In this work, we explore the potential of unsupervised learning through various forms of Variational Auto-Encoders (VAEs) \cite{maxkingma2013auto} in the context of biomarker research. We utilize an unsupervised representation learning approach in order to obtain objective tissue representations and explore to what extent learned representations form a complete description of the tissue by quantifying how well an input sample can be reconstructed from the latent representation (\textbf{I}),  are meaningful in the context of BE by investigating how well classifiers can predict tissue stage, taking only the latent representations as input (\textbf{II}), and are interpretable by exploring the generative capabilities of learned models (\textbf{III}).

In the context of representation learning, we refer to  interpretability as both the capability of generating images from latents (thus providing visual interpretation) and the ability to interpolate between learned representations. That is, in an ideal scenario, the latent space is organized in regions that correspond to different stages of BE, and interpolation would correspond to a smooth transitioning from healthy towards cancerous tissue via NDBE, LGD and HGD. It is known, however, that interpolation using VAEs suffers from latent-space distortion, in which case nonsensical images are generated along the trajectory \cite{chadebec2020geometryaware}. This can be avoided through geometric modeling of latent spaces (Sec. 1.1). 

In this paper we explore the importance of geometric latent space modeling by comparing hyperspherical VAEs \cite{davidson2018hyperspherical} to normal VAEs, and explore a recently proposed \textit{equivariant} variant \cite{vadgama2022kendall} that allows us to be insensitive to the arbitrary orientation in which tissue is imaged under a microscope \cite{lafarge2020orientation}. In particular, we address the objectives \textbf{I}-\textbf{III} through an extensive empirical study that compares different variants of (V)AEs: variational vs. non-variational; normal Euclidean vs. hyperspherical latent spaces; equivariant vs. equivariant architectures. Additionally, we solve the problem of hyperspherical VAEs being limited to small latent-space sizes, by proposing \textit{a new loss that turns hyperspherical autoencoders into generative models}.
\subsection{Related Work} 
In clinical settings, BE is diagnosed through endoscopic surveillance, where biopsies are taken from the esophagus lining and examined under a microscope. The Vienna criteria \cite{bolero} are used to classify the severity of dysplasia in BE, which is subdivided into Non-Dysplastic Barrett’s Esophagus (NDBE), Low-Grade Dysplasia (LGD), High-Grade Dysplasia (HGD), and an indefinite class for uncertain diagnoses. Pathologists use specific tissue grading features, such as clonality, surface maturation, glandular structure architecture, cytonuclear abnormalities, and inflammation, to make accurate classifications \cite{vanderwelBarrett}. 
Such morphological changes can be captured in the latent space of a variational autoencoder.

To mitigate the distortion issue of the original VAE, various VAEs utilizing non-Euclidean manifold have been proposed, such as Riemannian \cite{tosi2014metrics,arvanitidis2017latent,chen2018metrics,shao2018riemannian}, elliptic \cite{skopek2019mixed,bachmann2020constant,gu2018learning}, or hyperbolic \cite{davidson2018hyperspherical}. Notably, Davidson et al. \cite{davidson2018hyperspherical} proposes a spherical VAE framework (\svae) that operates on a hyperspherical latent space, allowing for more flexible and distortion-free representations. Furthermore, in the context of medical imaging, Lafarge et al. \cite{lafarge2020orientation} introduced an equivariant VAE model (\textit{SE}(2)-VAE), to tackle the issue of encoding irrelevant information, specifically orientation and translation. The SE(2)-VAE extends the traditional VAE with a group-convolutional neural network \cite{cohen2016group}, enabling the model to be invariant to arbitrary rotations and translations. 
Building upon these advancements, Vadgama et al. \cite{vadgama2022kendall} proposed the KS-VAE framework, combining a hyperspherical latent space and an orientation-disentangled group-convolutional network. 
% The KS-VAE framework by Vadgama et al. represents a promising solution for addressing both the distortion problem in VAEs and the encoding of irrelevant information, such as orientation, in medical image data. By combining the advantages of hyperspherical VAEs and equivariant VAEs, the KS-VAE provides a powerful approach to representation learning, particularly in the context of histopathological image data.

\section{Method}
\subsection{VAEs and Hyperspherical VAEs}
VAEs \cite{maxkingma2013auto} are powerful unsupervised learning models based on the assumption that data is generated via $x = D(z) + \epsilon$ with $\epsilon$ random noise and $D$ a so-called \emph{decoder}, that decodes the data content from a low-dimensional latent variable. It defines a conditional data distribution $p(x | z)$, the likelihood, which together with a prior distribution $p(z)$ on the latent space defines a distribution on the data space from which one can generate (sample) new data points. One is typically interested in obtaining the compressed latent variable $z$ for a given input $x$, which can probabilistically be done via the  posterior $p(z|x)$. However, the computation of the true posterior is typically intractable and one thus resorts to approximating it with a distribution $q(z|x)$ that is parameterized by an \emph{encoder} neural network $E$. Via the approximation, one does not directly maximize the (marginal) data-evidence, but instead the Evidence Lower Bound (ELBO) \cite{maxkingma2013auto}:
\begin{equation}
    \mathcal{L}_{\text{ELBO}} = \mathbb{E}_{q(z|x)}[\log p(x|z)] - \text{KL}(q(z|x) || p(z)),
\end{equation}
which consists of a reconstruction loss (measuring fidelity of reconstructed data) and the KL divergence between the approximate posterior and prior on $z$.

When the latent space is Euclidean, the approximate posterior $q(z|x)$ and prior $p(z)$ are usually normally distributed, with the parameters of $q(z|x)$ obtained via the encoder network, and those of $p(z)$ set as hyperparameters. 
% This leads to a distortion in the latent space, because the Gaussian prior tends to concentrate points in a cluster around the center of the distribution’s probability mass. This becomes especially problematic in case of multi-class data, as separate clusters in the latent space will also be pulled towards the origin, which makes them difficult to distinguish from each other \cite{davidson2018hyperspherical}. 
For hyperspherical latent spaces we need an equivalent of the normal distribution, which is given by the von Mises-Fisher (vMF) distribution:
\begin{align}
    q(\mathbf{z} \mid \mu, \kappa) & =\frac{\kappa^{m / 2-1}}{(2 \pi)^{m / 2} \mathcal{I}_{m / 2-1}(\kappa)} \exp \left(\kappa \mu^{T} \mathbf{z}\right),
\end{align}
where the mean $\mu$ is a unit vector $(\|\mu\| = 1)$, $\kappa$ is a precision parameter, and $\mathcal{I}_{n}(\kappa)$ denotes the modified Bessel function of the first kind at order $n = (m/2-1)$. For the special case of $\kappa=0$, the vMF represents a Uniform distribution on the $(m - 1)$-dimensional hypersphere $U\left(\mathcal{S}^{m-1}\right)$. The closed form for KL divergence term between a uniform distribution and vMF distribution is derived in \cite{davidson2018hyperspherical}. 

A key element of hyperspherical VAEs is that, due to the compactness of the latent space, it is possible to work with \textit{uniform priors that make sure that the entire latent space is utilized} and every $z \in S^{m-1}$ corresponds to a sensible data point $x$. In contrast, in Euclidean VAEs mass in the prior $p(z)$ is typically centered around the origin. Thus, only a fraction of the space is used, resulting in inefficiency and challenges in effectively modeling and separating clusters in the latent space. Hyperspherical models do not suffer from these limitations.

\subsection{Generative Hyperspherical Autoencoder Through a New Loss}
\begin{figure}[H]
      \centering
      \subfloat[\sae without spread loss.]{\includegraphics[width=0.35\textwidth]{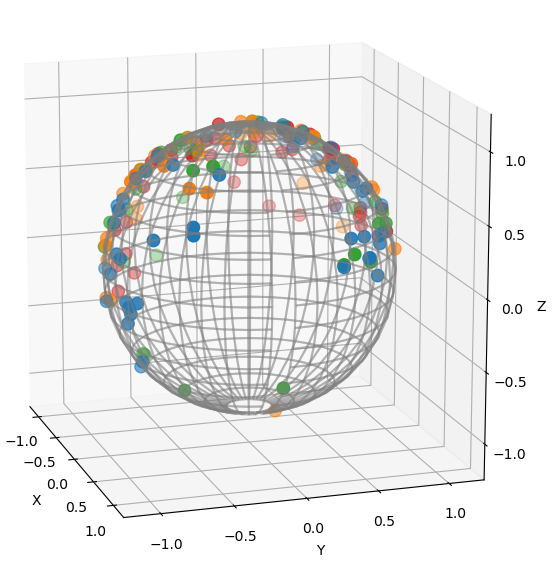}} \quad \quad \quad
      \subfloat[\sae with with spread loss.]{\includegraphics[width=0.35\textwidth]{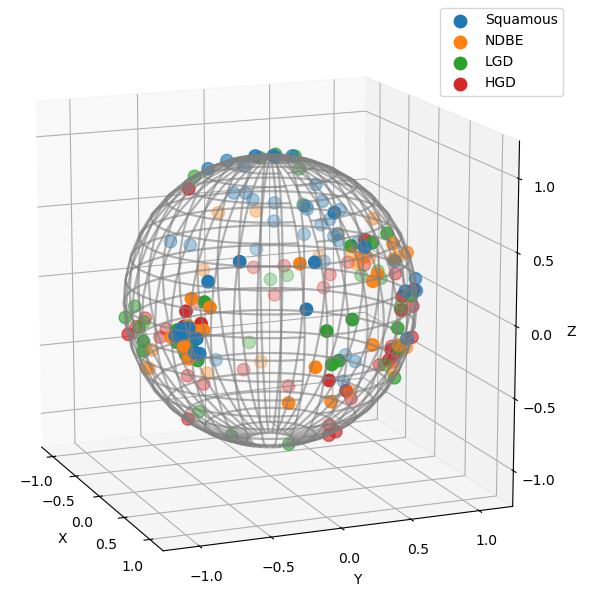} \label{fig:b}}
      \caption{Visualization of 3-D Latent Space for model \sae without and with spread loss. The same batch of 200 images was encoded by both models, and different image classes are visualized with different colored points. It can be observed that the points encoded by the model trained with spread loss cover a significantly larger area of the sphere.} \label{fig:spread}
    \end{figure}

Hyperspherical VAEs are known to be limited in generative capabilities when the dimensionality of the hypersphere $m$ becomes large, due to instability in sampling from the posterior vMF distributions \cite{davidson2018hyperspherical}. We solve this issue by leveraging the fact that, due to the uniform prior, the entire latent space is covered. That is, every $z \in S^{m-1}$ will equally likely generate a realistic sample of the learned data distribution. We then propose to avoid having to sample during training, by training an autoencoder (usually trained with only the reconstruction loss) with an additional loss that encourages a uniform coverage of data in the latent space which we call the \emph{spread loss}. The spread loss maximizes the distance between encoded data points in a batch via
\begin{equation}
\label{eq:spreadloss}
    L_{\text{spread}} = \sum^N_{i,j=1} -\mathbf{z}_i^T \mathbf{z}_j,
\end{equation}
where we note that maximizing the true distance $d(\mathbf{z}_i, \mathbf{z}_j)=\arccos(\mathbf{z}_i^T \mathbf{z}_j)$ is equal to minimizing (hence minus sign in \eqref{eq:spreadloss}) their inner products $\mathbf{z}_i^T \mathbf{z}_j$.  An example visualizing the effects of spread loss is shown in Figure \ref{fig:spread}.

 \subsection{Roto-Equivariant VAE and KS-VAE}
In addition to exploring different geometric latent spaces, we also investigate the idea of learning orientation-disentangled representations. Classic convolutional neural networks are not equivariant to rotation, causing the same image patches in different orientations to result in different learned representation vectors. Since orientation of scanned biopsies is arbitrary and the intrinsic properties remain unaltered by rotations, we want to learn rotation invariant representations. 

We modify (V)AEs to be rotation equivariant based on the method and code of \cite{bekkers2018roto}. We note that equivariance means that if the input rotates, the encoded latent transforms in a predictable manner via an action of the rotation group on the latent space. We follow the approach by Vadgama et al. \cite{vadgama2022kendall} which utilizes an equivariant encoder to obtain the latent representation $z$, together with a pose $\mathbf{R} \in SO(2)$ (a rotation matrix), which can be utilized to map $z$ to a canonical pose $z_0 = \rho(\mathbf{R}^{-1}) z$, with $\rho$ a representation of the rotation group acting on the latent space $S^{m-1}$. In their work it is shown that if the hyperspherical latent space is of dimension $(n-1)*2 - 1$, the latents $z$ can be interpreted and visualized as shapes/visual symbols that consist of $n$ two-dimensional landmarks in a Kendall shape space. The approach is similar to the equivariant VAEs developed by Lafarge et al. \cite{lafarge2020orientation}, except that \cite{vadgama2022kendall} canonicalizes latents $z$ via a predicted pose $\mathbf{R}$, and that our approach has a hyperspherical instead of Euclidean latent space.

\section{Experiments}

\subsection{Dataset}

We train the models on on a proprietary dataset retrieved from the Department of Pathology of the Amsterdam University Medical Centers and the LANS-panel (Dutch expert board of esophageal cancer). This dataset consists of digitized and annotated H\&E-stained endoscopic biopsies containing different BE progression stages. 
Additionally, we use the BOLERO dataset, which includes biopsies assessed by a panel of expert BE pathologists \cite{bolero}.
Combining these datasets, we have a total of 934 biopsies from 324 patients.
We use the BOLERO dataset as the test data. 
% This approach allows us to evaluate the generalization capabilities of our models and assess how well they capture different classes during training. 
We also reserve 10\% of the training set as a validation dataset.

\begin{figure}[h]
{\caption{Example of WSI and extracted image patches.}\label{fig:examplebiopsy}}
{\includegraphics[width=0.5\textwidth]{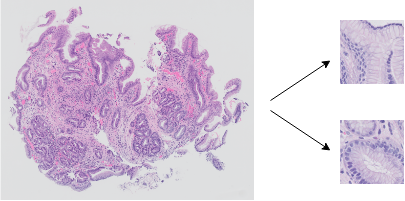}}
\end{figure}

The biopsies were digitized using a Philips Intellisite Ultrafast scanner and stored as Whole-Slide Images (WSIs), which are highly precise scans of glass slides containing multiple biopsies at various magnification levels. We preprocess the data by dividing the WSIs into smaller patches of size $64 \times 64$. See also Figure \ref{fig:examplebiopsy}. To ensure sufficient context, we choose a magnification level of $5\times$ and only include patches with a threshold of 50\% or more relevant tissue (squamous, NDBE, LGD, or HGD classes). 
Patch labels are computed based on pathologists' annotations in accompanying segmentation files, with the label determined by the dominant class within each patch. To balance the dataset and account for class imbalances, we stratify the dataset by selecting the 8,000 patches for each class, resulting in a balanced dataset of 32,000 patches.

\subsection{Experimental Setup}
We refer to hyperspherical and normal Euclidean VAEs as \svae and \textit{vanilla}-VAE respectively. In our experiments, we investigate representation learning models over three axes: 1) We compare hyperspherical to Euclidean latent spaces, 2) for each model we test both an equivariant (G-CNN) and non-equivariant (standard CNN) version, 3) we compare variational versus non-variational autoencoders. We employ the same architecture for all models, based on the work of Lafarge et al. \cite{lafarge2020orientation}. The encoder consists of three ConvNeXt blocks followed by max pooling, while the decoder mirrors this structure. The non-equivariant variational models generate parameters for the relevant posterior distribution, while the equivariant models also predict a pose per sample \cite{vadgama2022kendall}.
We train all models for 500 epochs with a batch size of 128, utilizing the Adam optimizer and MSE loss. We pad and normalize input images, excluding outer edges for equivariant models during reconstruction loss computation.

To fairly compare the models, we vary the latent dimension size and test sizes 3, 8, 16, 32, 64, 128, 256, and 512. As observed in previous research by Davidson et al. \cite{davidson2018hyperspherical} and confirmed in our experiments, high dimensions ($>$ 32) pose numerical instability for spherical models. To mitigate this instability, we introduce a minimum value of $\kappa$ (set at $\kappa=100$) which enables successful training of spherical autoencoders and VAEs up to a dimension size of 256. However, it is important to note that this approach limits the expressivity of the model, and this trade-off will be taken into account during result analysis. 

\section{Results} 
The experiments address three qualities of representation learning with the following questions: (\textbf{I}) are the learned representations complete (from a compression perspective); (\textbf{II}) are the learned representation semantically meaningful?; (\textbf{III}) what are the generative capabilities of each model?
\begin{table}[t]
\scriptsize
        \caption{Reconstruction losses on test dataset}
        \label{table:reconstructions}
        \centering
        \begin{tabular}{@{}lrrrrrrrr@{}}
        \toprule
              & \multicolumn{4}{c}{Normal}                                                                                                                                                      & \multicolumn{4}{c}{Spherical}                                                                                                                                                   \\ \cmidrule(l){2-9} 
              & \multicolumn{2}{c}{Non-Equivariant}                                                & \multicolumn{2}{c}{Equivariant}                                                            & \multicolumn{2}{c}{Non-Equivariant}                                                & \multicolumn{2}{c}{Equivariant}                                                            \\ \cmidrule(l){2-9} 
        $M$   & \multicolumn{1}{c}{\nvae} & \multicolumn{1}{c}{\nae} & \multicolumn{1}{c}{Eq. \nvae} & \multicolumn{1}{c}{Eq. \nae} & \multicolumn{1}{c}{\svae} & \multicolumn{1}{c}{\sae} & \multicolumn{1}{c}{Eq. \svae} & \multicolumn{1}{c}{Eq. \sae} \\ \midrule
        $3$   & 1895.56                                  & 2013.46                                 & -                                            & -                                           & 1930.60                                  & 2089.98                                 & -                                            & -                                           \\
        $8$   & 1807.06                                  & 1769.14                                 & 2103.47                                      & 2103.35                                     & 1707.14                                  & 1743.25                                 & 2103.66                                      & 2103.27                                     \\
        $16$  & 1635.89                                  & 1621.97                                 & 1290.60                                      & 1252.64                                     & 1563.19                                  & 1607.79                                 & 1303.07                                      & 1313.58                                     \\
        $32$  & 1435.60                                  & 1428.11                                 & 1161.32                                      & 1134.30                                     & 1389.64                                  & 1403.93                                 & 1142.24                                      & 1135.50                                     \\
        $64$  & 1260.85                                  & 1273.44                                 & 993.94                                       & 988.42                                      & 1250.45                                  & 1258.50                                 & 1009.90                                      & 992.39                                      \\
        $128$ & 1092.88                                  & 1113.56                                 & 857.40                                       & 853.79                                      & 1133.33                                  & 1104.18                                 & 902.82                                       & 853.75                                      \\
        $256$ & 904.53                                   & 935.09                                  & 710.22                                       & 706.50                                      & 1056.68                                  & 925.07                                  & 826.10                                       & 703.88                                      \\
        $512$ & 748.42                                   & 736.92                                  & 562.11                                       & 556.61                                      & -                                        & 727.06                                  & -                                            & 540.13                                      \\ \midrule
              & \multicolumn{1}{l}{}                     & \multicolumn{1}{l}{}                    & \multicolumn{1}{l}{}                         & \multicolumn{1}{l}{}                        & \multicolumn{1}{l}{}                     & \multicolumn{1}{l}{}                    & \multicolumn{1}{l}{}                         & \multicolumn{1}{l}{}                       
        \end{tabular}
    \end{table}
\begin{table}[t]
\caption{Classification Accuracy of Latent Representations on Test Dataset}
\scriptsize
\begin{tabular}{@{}lcccccccccc@{}}
\toprule
      & \multicolumn{6}{c}{Normal}                                                                      & \multicolumn{4}{c}{Spherical}                                         \\ \midrule
      & \multicolumn{2}{c}{Non-Equivariant} &               & Equivariant &          &                  & \multicolumn{2}{c}{Non-Equivariant} & \multicolumn{2}{c}{Equivariant} \\ \cmidrule(l){2-11} 
$M$   & \nvae             & \nae            & CNN           & Eq. \nvae   & Eq. \nae & Eq. CNN & \svae             & \sae            & Eq. \svae         & Eq. \sae    \\ \midrule
$3$   & 0.25              & 0.26            & 0.46          & -           & -        & \textit{-}       & 0.25              & 0.26            & -                 & -           \\
$8$   & 0.33              & 0.35            & \textit{0.48} & 0.34        & 0.17     & \textit{0.45}    & 0.36              & 0.33            & 0.23              & 0.27        \\
$16$  & 0.39              & 0.40            & \textit{0.47} & 0.39        & 0.42     & \textit{0.52}    & 0.40              & 0.34            & 0.34              & 0.30        \\
$32$  & 0.41              & 0.40            & \textit{0.46} & 0.47        & 0.46     & \textit{0.50}    & 0.41              & 0.31            & \textbf{0.49}     & 0.46        \\
$64$  & 0.45              & 0.40            & \textit{0.45} & 0.40        & 0.41     & \textit{0.54}    & 0.39              & 0.41            & 0.40              & 0.43        \\
$128$ & 0.42              & 0.42            & \textit{0.47} & 0.40        & 0.40     & \textit{0.51}    & 0.40              & 0.39            & 0.40              & 0.28        \\
$256$ & 0.42              & 0.42            & \textit{0.51} & 0.38        & 0.40     & \textit{0.50}    & 0.40              & 0.41            & 0.39              & 0.24        \\
$512$ & 0.38              & 0.36            & \textit{0.47} & 0.38        & 0.39     & \textit{0.51}    & -                 & 0.37            & -                 & 0.25        \\ \bottomrule
\end{tabular}
\label{table:classifications}
\end{table}    

Table \ref{table:reconstructions} addresses (\textbf{I}) following the idea that minimal information is lost if the decoder can reconstruct the input from the latent representation $z$. Here we observe the following: 1) increasing latent dimension size improves reconstruction fidelity; 2) the difference in variational vs non-variational autoencoders is small, but gets more pronounced in the hyperspherical showing that non-variational methods are preferred for compression; 3) equivariant methods have better reconstructions than non-equivariant ones; 4) hyperspherical latent space models outperform Euclidean ones.

Table \ref{table:classifications} characterizes the semantic meaning of learned representations (\textbf{II}) by testing how well we can train a classifier to categorize a given latent $z$ into each of the classes as given in Fig. \ref{fig:progression}. As a baseline, we trained a model with the default encoder architecture to directly predict class from the input patch. This should provide an upper bound on classification performance, as this model has access to all available (uncompressed) data to do the classification. The baseline accuracy (upper bound) is 0.51 for non-equivariant and 0.54 for equivariant CNN variants. From Table \ref{table:classifications} we make the following observations: 1) Latent dimensions 32 and 64 consistently achieve the highest accuracy across models; 2) hyperspherical VAEs overall give the best performance; 3) the performance of latent space classifiers is close to the upper bound, suggesting that semantic meaning is preserved by the encoders.

\begin{figure}[t]
       % \centering
      {\includegraphics[width=1\textwidth]{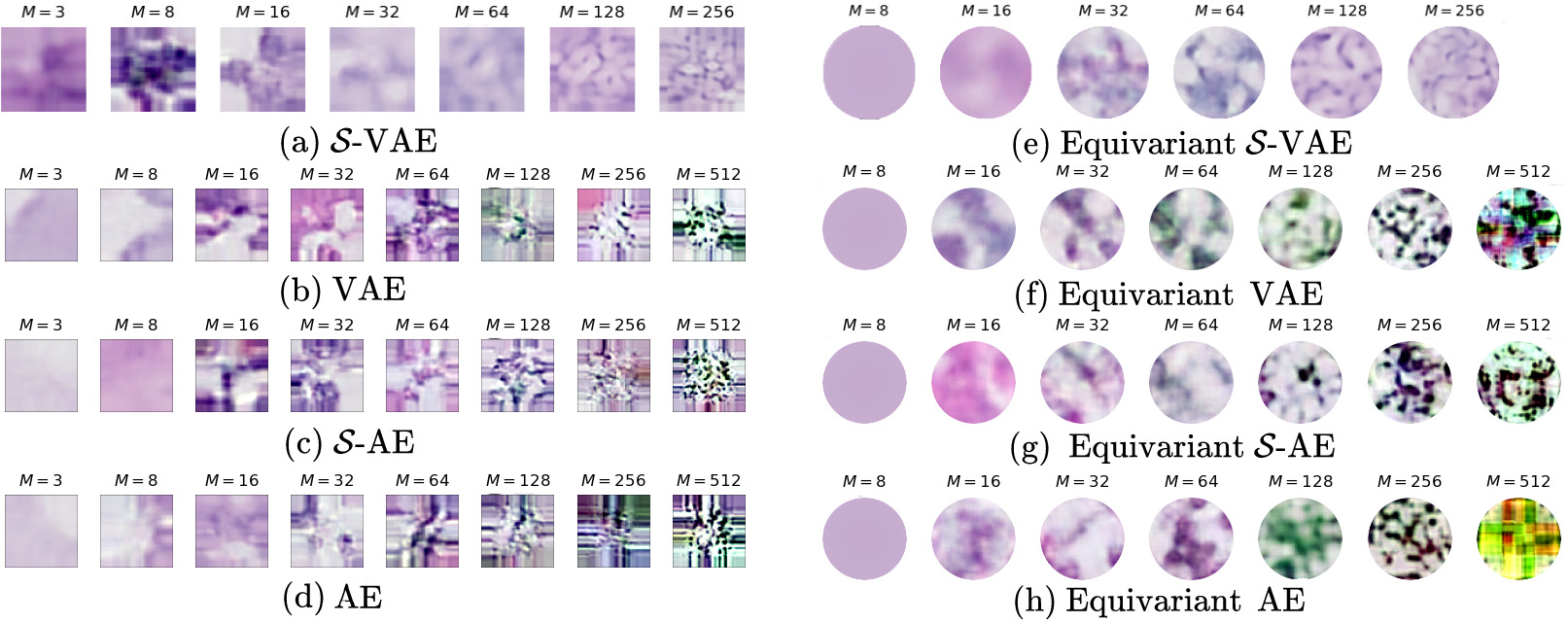}\vspace{-3mm}}
      \caption{Randomly generated images from all model types. Each column shows one sample from a model trained with a specific latent dimension size.}
      \label{fig:generations}
\end{figure}

To gain insight into the generative capabilities and visual interpretability (\textbf{III}) of the learned latent spaces, we sample vectors from random latent locations in all trained models and dimension sizes. Fig. \ref{fig:generations} showcases one sample per model and dimension size, demonstrating the general differences between the models. In terms of these generated images, a noticeable trend is the decrease in quality for higher dimension sizes across almost all models. Lower dimensions (3, 8, and 16) produce rough, blurry shapes with limited detail. However, in higher dimensions, images become less realistic, losing shapes and introducing colors not present in the original dataset. \textit{The only models capable of generating realistic images consistent with reconstructions in higher dimensions are the spherical VAE and its equivariant counterpart}. Among these models, equivariant \svae exhibits slightly more biopsy patch-like structures. \textit{However, even the best models generate images that are too blurry to consider them interpretable.}

\begin{figure}[t]
       % \centering
      {\includegraphics[width=1\textwidth]{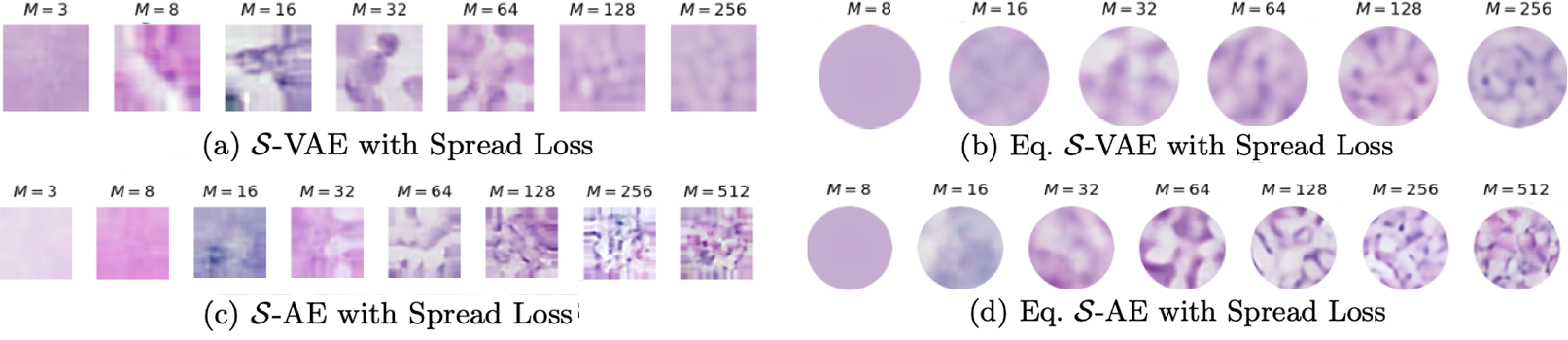}\vspace{-3mm}}
      \caption{Randomly image samples generated by \svae and \sae with spread Loss, for a range of latent dimension sizes.}
      \label{fig:generate_spread}
\end{figure}

Finally, to evaluate our novel \sae model and determine its potential as a generative model, we examine the effects of spread loss on the spherical autoencoder model by evaluating randomly generated images. From these images, shown in Fig. \ref{fig:generate_spread}, it becomes apparent that the introduction of spread loss to the spherical autoencoder substantially improves the quality of generated images. While generated images of autoencoder models previously looked unrealistic, with spread loss they resemble those generated by the variational models.

\section{Discussion}
Although the experiments provide important insights when it comes to design choices of (V)AEs, which we summarize in the conclusion, we also want to note in what sense the experiments are limited.
Firstly, the fidelity of reconstructions and generated samples are not yet at the level one hopes for in a context of interpretability. In comparison, the equivariant VAE of \cite{lafarge2020orientation}, whose neural network architecture we used as a baseline, gave high quality images of single nuclei. However, when scaling up to larger tissue areas, thus including clusters of cells, image quality degrades. We believe this is due to the large variability of cell positionings, their morphology and appearance. It seems that the image space is simply too diverse to be captured with the studied VAEs. The fact that the notion of \textit{equivariance} and \textit{hyperspherical latents} significantly improve image quality provides promising leads for future research.

Secondly, we explored capabilities to learn semantically meaningful representations via a classification analysis. Although the best methods came close to empirically found upper bounds on performance, the bounds themselves showed quite some room for improvement. I.e., ideally, the bound would be close to 100\% accuracy. The reasons we believe this is not achieved are two-fold. 1) We had to limit patch-size (and thus context window) in order to obtain reasonable image reconstructions with the (V)AEs. Going beyond this would further degrade reconstructed image quality, however, it would have given more context for the baseline classifier. 2) The labeling of tissue patches is a highly variable and subjective manual task. This is precisely the motivation for why we are investigating unsupervised learning methods. Nevertheless, the experiments show that neural networks can pick up on consistent semantic cues in an unsupervised manner.

\section{Conclusion}
In this study, we explored the application of several variants of VAE to learn tissue representations in an unsupervised manner, with the intent to develop tools that contribute to an objective understanding of the progression of Barrett's esophagus. Our contributions are threefold: 1) the experimental analysis of (V)AE variants showed the importance of \textit{equivariance} and \textit{hyperspherical latent space} modeling; 2) it showed the potential (latent representations can be semantically meaningful) and limitations (image generations show room for improvement) of generative unsupervised representation learning; and 3) we showed that one can train generative autoencoders in a non-variational setting without compromising on performance. Our novel spread loss allowed to train generative autoencoders without having to rely on a sampling, thereby circumventing the problem of limited latent space dimension of hyperspherical VAEs. 
Our study showed the stability of generative models with hyperspherical latent spaces and establishes a strong basis for further representation analysis via e.g., cluster analysis or interpolation experiments. We presented first steps towards a quantitative understanding of the latent space of esophageal tissue and how it could be organized along the axis of progression from healthy to cancerous tissue.

\vspace{0.5cm}
\noindent \textbf{Disclaimer:} This is the author-accepted version of the paper published in MICCAI 2023 Workshop proceedings, Lecture Notes in Computer Science (LNCS), Springer. The final version is available at: \url{https://doi.org/10.1007/978-3-031-45350-2_11}.

\bibliographystyle{unsrtnat}
\bibliography{mybibliography}

\end{document}